\documentclass[useAMS,usenatbib]{mn2e}

\def\lsim{\mathrel{\rlap{\lower4pt\hbox{\hskip1pt$\sim$}}
    \raise1pt\hbox{$<$}}}                
\def\gsim{\mathrel{\rlap{\lower4pt\hbox{\hskip1pt$\sim$}}
    \raise1pt\hbox{$>$}}}                

\begin{document}

\title[Gravitational Wave Background in the Quasi-Steady State Cosmology]{Gravitational Wave Background in the Quasi-Steady State Cosmology}

\author[J.V. Narlikar, S.V. Dhurandhar, R.G. Vishwakarma, S.R. Valluri and Sayantan Auddy]{J.V. Narlikar$^{1}$, S.V. Dhurandhar$^{1}$, R.G. Vishwakarma$^{2}$, S.R. Valluri$^{3,4}$ and S. Auddy$^{3}$ \\
 $^1$ Inter-University Centre for Astronomy and Astrophysics, Post Bag 4, Ganeshkhind, Pune 411 007, India\\
$^2$Unidad Academica de Matematicas, Universidad Autonoma de Zacatecas, C.P. 98068, Zacatecas, Mexico \\
$^3$Department of Physics and Astronomy, University of Western Ontario, London ON N6A 3K7, Canada\\
$^4$ Kings University College, University of Western Ontario, London ON N6A 2M3, Canada}
\date{}
\maketitle

\begin{abstract} 

This paper calculates the expected gravitational wave background (GWB) in the quasi-steady
state cosmology (QSSC).  The principal sources of gravitational waves in the QSSC are the
minicreation events (MCE).  With suitable assumptions the GWB can be computed both 
numerically and with analytical methods.  It is argued that the GWB in QSSC differs from that
predicted for the standard cosmology and a future technology of detectors will be able to 
decide between the two predictions.  We also derive a formula for the flux density of a typical 
extragalactic source of gravitational waves.     

\end{abstract}

\section{Introduction}

There have been several multiwavelength tests of cosmological models involving electromagnetic radiation. On an independent note it is hoped that additional tests may be eventually forthcoming as the technology of detecting gravitational radiation improves. In a previous paper (Sarmah, et al.,2006, hereafter referred to as Paper I), it was shown that a new type of source of gravitational radiation is suggested by the quasi-steady state cosmology (QSSC in  brief hereafter)in the form of a typical {\it, mini-creation event}. It was argued that these events may just be detectable by  the next generation detectors. 

In this paper we will try to estimate the radiation background produced by gravitational waves emitted by  minicreation events (MCEs in brief). This result should be of interest because of its comparison with the prediction of a relic gravitational wave background in the standard (big bang) model of the universe. We will make such a comparison and suggest the features that observations of the gravitational wave background (GWB in brief) may look for.

In the following sections we begin with a discussion of what QSSC is and how it is dynamically driven by the MCEs. In Section 3, we derive the expected GWB arising from the MCEs. In Section 4, we  attempt a comparison of our derived result with the standard model as well as with observations.   
In the concluding section we will highlight the importance of such a calculation in our quest for the right cosmology.

\section{The quasi-steady state cosmology}

\subsection{The mathematical model}

The QSSC model was first proposed by Hoyle, Burbidge and Narlikar, to be collectively referred to as HBN, in 1993. Their original paper (HBN, 1993) was followed by several others in the following years (see, for example, 1994a, 1994b) including a technical monograph (HBN, 2000), which gives a comprehensive account of the QSSC model. The cosmology uses the Machian theory of gravity by Hoyle and Narlikar (1964,1966) modified to include creation of matter. The creation terms are essentially described by a negative energy scalar field. Additionally there is a cosmological term of the form similar to the $\lambda$ term in relativity, except that it has the opposite sign.  The field equations are given by

\begin{equation}
R_{ik}-\frac{1}{2}g_{ik}R+ \lambda g_{ik}= - \frac{8\pi G}{c^4}[T_{ik} - f(C_i C_k -\frac{1}{4}g_{ik}C^lC_l)]
\end{equation}

\noindent where $f$ is a coupling constant while $\lambda$ is the cosmological constant.  $\lambda$, however, has the opposite sign (negative) in this cosmology.  

The QSSC model arises as a combination of two types of solutions of the above equations. The cosmological solutions after using the symmetries of the Weyl Postulate and the Cosmological Principle, are described by the Robertson-Walker line element with vanishing curvature parameter $k$ :

\begin{equation}
ds^2 = c^2 dt^2 - S^2 (t) [dr^2 + r^2 (d \theta^2 + {\rm sin}^2 \theta d \phi^2)].  
\end{equation}

The function $S(t)$ describes the scale factor of the universe.   

The two types of solutions arise depending on whether matter is being created or not. The combination of the two alternatives is represented by the solution for the scale factor

 \begin{equation}
S(t) = {\rm exp} (t/P) \bigg[1 + \eta {\rm ~cos} \frac{2\pi \tau}{Q}\bigg]. 
\end{equation}

Here the part  exp$(t/P)$ represents the creative part which recalls the classical steady state model of Bondi and Gold (1948) and Hoyle (1948). If there were no other term, this would describe a steady state universe with a constant rate of expansion given by Hubble's constant equal to 1/$P$ as measured by the time coordinate $t$. The second term in the scale factor represents oscillations of period $Q$ as measured by the time coordinate $\tau$. $\tau$ is a nonlinear function of $t$ but it is more or less proportional to $t$ except close to the minimum values of $S$. Henceforth we will take $\tau = t$. Accordingly the scale factor is modified from (3) above to

\begin{equation}
S(t) = {\rm exp} (t/P) \bigg[1 + \eta {\rm cos} \frac{2\pi t}{Q}\bigg]. 
\end{equation}

Also,we have $P>>Q$ and the constant $\eta$ lies in the open interval (0,1) so that the scale factor is $nonsingular$. The details of these QSSC solutions may be found in the paper by Sachs, et al.(1996). With reference to the epoch $t = 0$, we note that the maxima of $S$ lie at epochs equalling $Q$, 2$Q$, 3$Q$,...in the future values of $t$ and 0,$-Q$, $-2Q$,...in the past epochs.  In the same way,
the minima lie at $t$-values $0.5Q$, $1.5Q$,... in the future and $-0.5Q$, $-1.5Q$...in the past.

In such a model with time-axis extending from past infinity to future infinity, each cycle is physically the same as all others. This is because, matter creation occurs selectively near the minima of the scale factor for a reason which will be given shortly. Since because of the exponential term in the scale factor, the density of the universe drops off by a factor exp $(-3Q/P)$ between two successive minima, the creation of matter occurs to compensate for the drop. (See Sachs, et al. op.cit.)This is the reason for calling the cosmology ``Quasi-Steady". 

In this scenario, where do we place ourselves? Since all cycles are alike, we can choose any! Without loss of generality we place our present epoch at a $t$-value between 0.5$Q$ and $Q$ (note now the minimum $0.5 Q$ lies in the past). Denoting the present epoch by $t_0$ we determine it by using the observed value of Hubble's constant:

\begin{equation}
H(t_0) = \frac{\dot{S}}{S}\mid_{t_0}
\end{equation}  

Relations like these help determine the values of the parameters of the model, namely, $P$, $Q$, $\eta$ and $t_0$.
HBN (2000) gives details of how this can be done. {\it The important point is that given four constraints to determine these parameters, the QSSC is vulnerable to any more observational constraints, just as the old steady state theory was constrained by observations.} 

To fix ideas, we will follow the analysis given in Chapter 16 (pages 199 - 201) of reference
(HBN, 2000) to fix the numerical values of the parameters as under :  

$$P = 20Q, ~~Q= 42 \times 10^9 {\rm ~yrs}, ~~ \eta=0.85 $$

$${\rm Redshift ~of ~ last ~minimum} = z_m = 8 $$

$${\rm Present~epoch~} t_0 = \frac{1}{2}Q + 0.3 Q $$

Note that the maximum redshift expected in this model with the above parametric values is
8.  As explained in the discussion above, slightly different values of the parameters will give
slightly different answers for $t_0$ and $z_m$.  The QSSC authors have argued there that
available data may be used to give values to these numbers and then the theory be used 
for testing. Accordingly we adopt these values for our estimates of gravitational wave 
background.   

\subsection{The minicreation events}

We now come to the creation process itself. We confine ourselves to a brief description, referring to HBN (2000), Chapter 18 for details. The creation of matter in this  cosmology is in the form of the Planck particle with mass, 

\begin{equation}
m_{\rm Pl} = \bigg(\frac{3 \hbar c}{4\pi G}\bigg)^{1/2}
\end{equation}

Indeed, given the fundamental constants $G$, $c$ and $\hbar$,this is the only combination with dimensions of mass.  Since the field equations tell us that the condition for creation is the equality

\begin{equation}
C_i = p_i ({\rm momentum}) ; p_i p^i = m^2_{\rm Pl} c^2 
\end{equation}

\noindent describing a balance between the energy-momentum of $m_{\rm Pl}$ created and the negative energy of the $C$-field present, there is no violation of the conservation law for energy. The C-field idea, first introduced by Maurice Pryce (private communication) in 1961 was extensively used by Hoyle and Narlikar (see for example, 1962,1963,1964a,1964b,1966).  Although considered unphysical in the sixties, it has resurfaced as phantom fields (Sami 2004) today.

The creation condition is in general not possible to satisfy in view of the large mass of the Planck particle. It can, however, be satisfied in a strong gravity environment. If we consider the Schwarzschild type metric, the $C$-field strength shoots up as one goes closer to the Schwarzschild radius as shown below:

\begin{equation}
C^i C_i \propto \bigg(1 - \frac{2GM}{c^2R}\bigg)^{-1}
\end{equation}

In general relativity a black hole forms through gravitational collapse of a massive object. If $C$-field is present, there is a bounce of the collapsing object just outside the Schwarzschild radius. This is where a condition for creation of new matter is possible. Since, as seen in (7), creation of matter has to be balanced by $C$-field, we also get the $C$-field created. And, because of its negative energy, it creates, locally, a {\it repulsion} force that drives away the created mass. Thus we have a finite, nonsingular event resulting in explosive creation of matter. This is called a {\it mini-creation event}. We shall henceforth refer to it as MCE.

A more likely form of MCE considered in Paper-I arose from a Kerr-type spinning object. Such an object would result in ejection along two oppositely directed jets along the poles as there is least resistance to ejected particles in moving away.

\subsection{Observational tests}

Several observational tests have been applied to QSSC such as the redshift magnitude relation, radio source count, creation of light nuclei, relic radiation peaking at microwave wavelengths, formation of large scale structure, etc. Details can be found in HBN (2000) and later papers of Narlikar et al. (2002),(2003), Vishwakarma and Narlikar (2010). 

Additionally, QSSC has also suggested a few potential tests that distinguish it from the standard model. These include the finding of very old (age $\sim$ 20 Gyr) stars, discovery of blueshifted galaxies beyond 27th magnitude, baryonic matter density exceeding the limit permitted by big bang etc.  

To this last category we now wish to add the input provided by observations of the gravitational wave background. We will next show how we may compute such a background in a form that can be compared to the result expected from standard cosmology via inflation. As and when technology progresses to a level that one can actually carry out background measurements, it is useful to have theoretical predictions ready.
 
We wish to clarify here that this paper is limited to the topic of gravitational waves only in the 
role they might play in testing cosmological models like the standard model and the QSSC.  As
indicated at the beginning of this subsection, some work has been done using different 
wavelengths of electromagnetic radiation to constrain the parameters specifying the QSSC.  
The main theme of this paper is not concerned with the findings of those tests, although we
expect a review of all such observational tests will eventually decide on the viability of QSSC.  

In this connection, the use of the Alcock-Paczynski test by Lopez-Corredoira (2014) to exclude
certain models including the QSSC is a recent addition to these other observational tests. Although the probability of the QSSC model is low, 2 $\%$ is not low enough to definitely discard the QSSC, and that the statistics with the data analysed may be sensitive to methods used to disentangle redshift distortions and geometric cosmological distortions 
(private communication \footnote{Authors are grateful to Lopez-Corredoira for pointing this out.}).   A paper
reviewing tests like these along with the others mentioned earlier will indeed be timely and we 
plan to take it up as a separate exercise.  

\section{Computation of gravity wave background in the QSSC}

We will estimate the total contribution to the gravitational wave background in the QSSC, on the assumption that the background is built up from contributions made by all Mini-Creation Events.  Thus we will include contributions of $all$ MCEs from $all$ past cycles of the cosmological model. To this end we first estimate the gravitational waves emitted by a typical MCE.

\subsection{Gravitational waves from a typical MCE}
 In Paper I there is an extensive discussion of this topic and we can do no better than draw on the results obtained there. 
As described in the preceding section, the MCE may be visualized as a twin jet event which ejects newly created   matter in opposite directions. Let $\dot{M}$ denote the rate of creation of matter in the MCE and suppose that the created matter is moving in the two jet directions with speed $u$. In Paper I it was shown that the radiation reaction does not slow down the source significantly.  

The formula (18) in Paper I gives the rate of emission of such an MCE:

\begin{eqnarray}
L_{GW} &=& \frac{c^3}{16 \pi G} \alpha \bigg(\frac{4G\dot{M}u^2}{c^4 R}\bigg)^2 \cdot 4 \pi R^2 \nonumber \\
&=& \frac{4G\dot{M}^2 u^4\alpha}{c^5}
\end{eqnarray}

\noindent where $\alpha$ is a dimensionless constant of order unity and $L_{GW}$ is the luminosity of the MCE integrated over all frequencies.   

To fix ideas we will assume that a typical MCE emits newly created matter at the rate of 200 solar masses per second and take $u = \beta c$. Formulae (21) and (22) of Paper I give the Fourier transforms of the gravitational wave amplitude for the two polarizations as    
   
\begin{equation}
\tilde{h}_+ (\nu) = \frac{\dot{M}Gu^2}{\pi^2 c^4 R}. \nu^{-2} \cdot {\rm ~sin}^2 \epsilon {\rm ~cos~} 2 \Psi
\label{polplus}
\end{equation}

\begin{equation}
\tilde{h}_\times (\nu) = \frac{\dot{M}Gu^2}{\pi^2 c^4 R}. \nu^{-2} \cdot {\rm ~sin}^2 \epsilon {\rm ~sin~} 2 \Psi
\label{polcross}
\end{equation}

These formulae are based on angular spherical coordinates $\epsilon$ and $\psi$ for the direction of the jet. Although the frequency $\nu$ seems to cause infrared divergence, as was explained in Paper I, there is an effective cut off because of bounded timescales of the sources.

Although, the assumed geometry of a typical MCE was rather special, we will allow for variations in it and the infrared divergence may be softened by the frequency dependence being just ${\nu}^{-1}$ over a finite range ($\nu_{min},\nu_{max}$).

So the emission rate of an MCE may be taken as

\begin{equation}
L_{GW} (\nu) d\nu = \frac{4G\dot{M}^2u^4 \alpha}{c^5 \nu^2} Kd\nu
\end{equation}

\noindent where $\nu_{min} < \nu_{max}$ and $K$ is chosen so that 

\begin{equation}
K \int^{\nu_{max}}_{\nu_{min}} \frac{d\nu}{\nu^2}=1
\end{equation}

\noindent Thus, when $\nu_{max}\gg \nu_{min}$, we have $K = \nu_{min}$.  We will 
assume this to be the case.  

Finally, we need to feed in information of the number densities of the MCEs and their creation rates.  We relate this information to the dynamics of the QSSC in the following way.

Consider two successive minima of scale factors, separated by the period $Q$. The density of matter at the start of the cycle is denoted by $\rho$, say. Because of the secular expansion factor in $S$, the matter density would drop by the factor $\exp (-3Q/P)$
at the next minimum epoch. However, the creation of new matter mainly through the MCE activity, would restore the density to
its previous value at the start of the cycle. This tells us that the density of matter created will be

\begin{equation}
\rho_{cr} = \rho \bigg[1 - \exp \bigg(- \frac{3Q}{P}\bigg)\bigg].
\end{equation}

\noindent  We will estimate this figure by taking the present density $\rho_0$ as 

\begin{equation}
\rho_0 = \frac{3H^2_0}{8\pi G}, 
\end{equation}

\noindent $H_0$ being the present Hubble constant.  

The density at the minimum epoch will therefore be 

\begin{eqnarray}
\rho &=& \rho_0 \frac{S(t_0)\mid^3}{S(Q/2 )^3} \nonumber \\      
&=& \rho_0 \bigg(\frac{1 + \eta {\rm ~cos~} \frac{2\pi t_0}{Q}}{1-\eta}\bigg)^3  \exp \left [\frac{3 (t_0 - \frac{1}{2}Q)}{P} \right] \,.
\label{dnsty}
\end{eqnarray}

Note that given the QSSC parameters $t_0$, $P$, $Q$ and $\eta$ we have $\rho_{cr}$ fully determined.                       

This has to be equated to the matter created per unit volume through the MCE activity. If each MCE generates matter at the rate of $q$ solar masses per second and the activity lasts for $T$ seconds, then we have the number of MCEs per unit volume as

\begin{equation}
N = \frac{\rho_{cr}}{qTM_\odot}.  
\end{equation}

\noindent Using (14) and (17) we get 

\begin{equation}
N = \rho \bigg[1 - {\rm exp} \bigg(- \frac{3Q}{P}\bigg)\bigg].\{qTM_\odot \}^{-1} \,.
\label{ndnsty}
\end{equation}  

\subsection{Gravitational wave background}

It is convenient to begin with the formula used commonly by optical astronomers when evaluating the dimming of a source of radiation observed from far away. The formula is given in any standard text of cosmology; see for example, the book by Narlikar (2002). Given that a source with redshift $z$ has luminosity $L$ distributed as a spectrum $F(\nu)$d$\nu$ so that
                     
\begin{equation}
\int^\infty_0 F(\nu) d\nu = 1, 
\end{equation}

\noindent then the flux of radiation crossing unit normal area at the 
observer over the frequency band $[\nu,\nu+d\nu]$ will be

\begin{equation}
\zeta (\nu) = \frac{LF (\nu \cdot \overline{1+z})}{4\pi r^2 S^2_0 (1+z)}. 
\end{equation}

We have to sum these expressions evaluated at all past epochs of minimum $S(t)$.  As we saw before, these occur at epochs

\begin{equation}
\frac{1}{2}Q, -\frac{1}{2}Q, -\frac{3}{2}Q, \cdots .
\end{equation}

\noindent i.e., at 

\begin{equation}
t_n = - \bigg(n - \frac{1}{2}\bigg) Q 
\end{equation}

\noindent for $n = 0, 1, 2, \cdots .$

Redshifts of these epochs are respectively

\begin{eqnarray}
z_n &=& \frac{S(t_0)}{S(t_n)}-1 \nonumber \\
& = & {\rm ~exp~}\frac{t_0 - t_n}{P}\cdot \frac{1 + \eta {~cos~} \frac{2\pi t_0}{Q}}{1-\eta} -1.
\end{eqnarray}

We now carry out essentially an Olbers-type calculation which in earlier times led to the 
well-known Olbers paradox.  Only, we use here gravitational wave background instead of
the optical background.  

Taking ourselves as located at $r=0$ with the present epoch $t_0$, we find that if a 
radiation pulse, emitted by an MCE at a minimum epoch $t_n$ is to reach us here and now, 
its distance from us has to be 

\begin{equation}
r_n = \int^{t_0}_{t_n} \frac{cdt}{S(t)}. 
\end{equation} 

\noindent Let us suppose that the creation activity of MCEs lasted for a short period after 
the minimum epoch. Although we are assuming that the MCEs occur during a short period, {\it they occur all over the universe}. Thus the GWs we receive come to us from different radii at different times forming a continuous wave background. Now we proceed with the calculation.  

Suppose a thin shell of radial thickness

\begin{equation}
\Delta = (1 + z_n)^{-1} {\rm ~second}
\end{equation} 

\noindent is sending gravitational radiation to $r=0$, reaching the observer (i.e. ourselves) there
lasting for a period of 1 second of our time.  The volume of this shell will be 

\begin{equation}
\Delta V_n = 4 \pi r^2_n (1+z_n)^{-1}S^2_0 c. 
\end{equation} 

Although non-Euclidean geometry might modify this formula somewhat, we will proceed with 
the above Euclidean formulation since the differences between geometries are unlikely 
to be significant.  

Hence the number of MCEs contributing to GWB in our neighbourhood is given by 
$\Delta V_n \cdot N$. The flux density contributed by each MCE over the range of frequencies
($\nu$ , $\nu$ + $d\nu$) is $\phi_n(\nu ) d\nu$ where 

\begin{equation}
\phi_n(\nu ) = \frac{\alpha G\dot{M}^2 u^4 \nu_{\rm min}}{\pi r^2_n c^5 \nu^2 (1+z_n)^3 S^2_0}.
\label{flux_1}
\end{equation} 

\noindent The suffix $n$ indicates that $\phi_n(\nu)$ originates in the creation process just 
after the $n^{\rm th}$ minimum [$n$ = 0, 1, 2, ...]. Summing over all $n$ gives the total 
contribution of the past MCEs as $B(\nu )d\nu$ where 

\begin{equation}
B(\nu ) = \sum^\infty_{n=0} \frac{4 \alpha G \dot{M}^2 u^4 \nu_{\rm min} N}{c^4\nu^2(1+z_n)^4} 
\end{equation} 

To fix ideas we substitute typical QSSC values, $P = 20Q$, $Q$ = 42Gyr and $\eta = 0.85$ and also substitute the value of $N$ from Eq. (\ref{ndnsty}) and the density at minimum epoch from Eq. (\ref{dnsty}).  This
gives us the following equation for the GW flux: 

\begin{eqnarray} 
B(\nu ) & = & \frac{4 \alpha G\dot{M}^2 u^4 \nu_{\rm min}}{c^4 \nu^2} \frac{\rho_0}{q T M_\odot} \, \nonumber \\
& \times & \bigg(\frac{0.15}{1 + 0.85 (\cos 1.6 \pi)}\bigg) e^{-0.015} \left( 1 - e^{-3/20} \right) \, \nonumber \\ 
& \times & \sum^\infty_{n=0} {\rm ~exp~} \bigg(- \frac{n}{5}\bigg) \,.
\label{flux} 
\end{eqnarray}

\noindent The series in Eq. (\ref{flux}) can be easily summed and yields,

\begin{equation}
\sum^\infty_{n=0} {\rm ~exp~} (-\frac{n}{5}) \simeq 5.52 \,.
\end{equation}
 
\noindent We may follow the model proposed in Paper I and take $u = \cdot 8c$ and $T = 1000$ sec.  The value 
of $B(\nu)$ integrated over all $\nu$ gives the total GWB as,

\begin{equation}
 \int_{\nu_{\rm min}}^{\infty} d \nu ~B (\nu) \sim 6.8 \times 10^{-5} \left (\frac{T}{1000 ~{\rm sec}} \right )^{-1} ~{\rm ergs}~{\rm cm}^{-2}~{\rm s}^{-1} \,.
\end{equation} 

\noindent In general, though, only the first few terms of the series with the values $n$ = 0, 1, 2, ... will contribute significantly to the sum. 
\par

The GW flux per unit frequency is,
\begin{equation}
B (\nu) \sim 6.8 \times 10^{-5} ~\frac{\nu_{\rm min}}{\nu^2} \left (\frac{T}{1000 ~{\rm sec}} \right )^{-1} ~{\rm ergs}~{\rm cm}^{-2}~{\rm s}^{-1} {\rm Hz}^{-1} \,.
\label{dflux}
\end{equation} 

\noindent If we take $\nu_{\rm min} = T^{-1}$ for the MCE source which expands for $T$ secs, and convert the flux from  Eq. (\ref{dflux}) into energy density of gravitational waves we get in 
comparison with the closure density - $\Omega \cong 10^{-29}{\rm g~cm}^{-3}$ - the expected energy density
of gravitational waves $\Omega_{GW} (\nu)$ as, 

\begin{equation}
\Omega_{GW} (\nu) \cong 1.4 \times 10^{-12} \left(\frac{\nu}{10~ {\rm Hz}} \right)^{-2} \left (\frac{T}{1000 ~{\rm sec}} \right )^{-2} \,.
\end{equation} 

The Einstein Telescope is believed to have the sensitivity to be just close to observing $\Omega_{GW}$ at $10Hz$ (vide Sathyaprakash and Schutz 2009). However, these are just order of magnitude estimates. If we had taken $T \lsim 160$ sec., then $\Omega_{GW} (\nu) \simeq 5 \times 10^{-11}$ which would be in the detection ballpark of the Einstein telescope. Any such detection would constrain the QSSC model. Integration over $\nu$ gives $\Omega_{GWB}\approx 1.4 \times 10^{-8}$.  This is an indicative figure since at present we have very little information about such sources.  

\subsection{GWs from a single MCE} 

Apart from the gravitational wave background derived above, it is useful to derive an 
expression for the flux of radiation received from an MCE source at the $n$th epoch.  
Eq. (\ref{flux_1}) gives the formal expression for this flux; but its simplification is of 
interest (and possible future use) when gravitational wave detectors are able to pick 
out individual sources.  

Starting with Eq. (\ref{flux_1}) and putting in QSSC values we get,

\begin{equation}
\phi_n(\nu ) = \frac{\alpha G \dot{M}^2 u^4 \nu_{\rm min}}{c^7 \nu^2 Q^2} I_n \,,
\end{equation}
where,
\begin{eqnarray}
I_n &=&  \frac{e^{-0.125}(0.15)^3}{\pi \{1+0.85 {\rm ~cos~}(1.6\pi)\}^5}{\rm ~exp~} \bigg(-\frac{3n}{20}\bigg)\nonumber \\
&\times&\bigg[\int^{0.8}_{\frac{1}{2}-n} \frac{e^ {-\tau /20}d\tau}{1+ 0.85 {\rm~cos~} (2\pi \tau) }\bigg]^{-2} \,.
\label{intgrl}
\end{eqnarray} 
We can compute $I_n$ numerically for a few typical values, say, $n = 0, 1, 5, 10, ...$. They are mentioned below in Table I. 

\begin{center}

Table 1 \\
\begin{tabular}{ll} \hline 
n  & ~~~ $I_n$ \\ \hline 
0 & 0.000459557\\
1 & 0.0000348655 \\
5 & $1.08893 \times 10^{-6}$ \\
10 & $1.06947 \times 10^{-7}$ \\
50 & $9.49022 \times 10^{-13}$ \\
100 & $3.03132 \times 10^{-18}$ \\ \hline 
\end{tabular}

\end{center}

We now show how an analytical solution can be obtained for the above expression, especially for the integral.  
We expand the denominator of the integrand in Taylor series in even and odd powers of $-\eta {\rm ~cos~} 2 \pi \tau $ and rewrite the integral as:
\begin{equation}
\sum^\infty_{m=0}\int^{0.8}_{\frac{1}{2}-n} e^{-A \tau} \left[(-\eta {\rm ~cos~} b \tau )^{2m}+(-\eta {\rm ~cos~} b \tau )^{2m+1}\right] d\tau
\label{analytic}
\end{equation}
We can separately solve for the even and the odd parts and then sum over $m$ to obtain the  analytic value of the integral (Prudnikov et al., 1986). For convenience we firstly give the integrals for the even and odd parts and then go back to the summations.
\\
Case 1 : The even part of the integral is:
\begin{eqnarray}
\int e^{-A\tau } ({\rm ~cos~} b \tau )^{2m} = -{2m\choose m} \frac{e^{-A\tau }}{2^{2n}A}+
\frac{e^{-A\tau }}{2^{2n-1}} \times \\ \nonumber \sum^m_{k=1} {2m\choose m-k}\frac{-A{\rm ~cos~} 2 k b \tau +2 b k {\rm ~sin~} 2 k b \tau}{A^2 +4 b^2 k^2}
\end{eqnarray}
Case 2 : The odd part of the integral 

\begin{eqnarray}
\int e^{-A\tau } ( {\rm ~cos~} b \tau )^{2m+1}= \frac{e^{-A\tau }}{2^{2n}}\sum^m_{k=0} \frac{1}{A^2+b^2 (2k+1)^2} \\ \nonumber {2m+1\choose m-k} (-A {\rm ~cos~} (2k+1) b \tau + (2k+1)b {\rm ~sin~} (2k+1) b \tau)
\end{eqnarray}
where $b=2 \pi$ and $A =1/20$.\\
It is interesting to observe that these integrals can be expressed in terms of the polylogarithm functions (Molli et al. 2011, Lewin 1981).
On substituting the appropriate limits in the above expressions followed by summation over $m$ and then using our analytic expression of  Eq. (\ref{analytic}) in Eq. (\ref{intgrl}), we obtain the same numerical values of the integrals $I_n$ for various values of $n$ given in Table I and so also for $\phi_n(\nu )$.  
\par
Of more interest is the estimate of the signal-to-noise ratio (SNR) of a single MCE occurring at the last minimum epoch $t = 0.5 Q$. The gravitational wave signal from a MCE is a linear combination of the two polarisation amplitudes 
$\tilde{h}_+ (\nu)$ and $\tilde{h}_{\times} (\nu)$ given by Eqs. (\ref{polplus}) and (\ref{polcross}) involving orientation factors. Averaging over the orientations, the average SNR $\rho$ is given by the equation:
\begin{equation}
\rho = 2 \alpha \frac{\dot{M} G u^2}{\pi^2 c^4 R} \left [\int_{\nu_{\min}}^{\nu_{\max}} \frac{d \nu}{\nu^4 S_h (\nu)} \right ]^{1/2} \,,
\end{equation} 
where $\alpha \lsim 1$ is the average orientation factor and $S_h (\nu)$ is the one sided PSD of the ET-D configuration given by Hild et al (2011). Taking ${\dot M} \sim 200 M_{\odot}$ per sec, $u \sim 0.8 c$ and  
\begin{equation}
R = c \int_{0.5 Q}^{0.8 Q} \frac{e^{- t / 20 Q}~dt}{1 + 0.85 \cos (2 \pi t/Q)} \sim 10 ~{\rm Gpc} \,,
\end{equation}
we estimate an average SNR of 2.3. This value is in the same ballpark as that obtained by Marassi et al (2011) for other sources of the stochastic gravitational wave background.

\subsection{Some Caveats} 

The expression for $B(\nu )$ derived above may, however, lead to a gross under-estimate 
of the gravity wave background.  For, the typical MCE chosen to contribute to the background, 
as taken over Paper I, is a powerful one.  Indeed, in Paper I we were interested in the detection
of an MCE by detector technology of foreseeable future.  In general, the QSSC expects creation
events of various strengths.  The bulk of them will be numerous but weak and so may not be
individually detected.  A powerful MCE of the kind chosen to give $B(\nu )$ will have 
emission of newly created matter at the rate of $200 M_\odot$ per second, travelling outwards
at velocity $u = 0.8c$.  If instead we had chosen the typical ejector to be working at $M_\odot s^{-1}$ and ejecting matter at speed $u = 0.99c$, then a calculation similar to (17) would give 
$N \propto (qTM_\odot )^{-1}$ a value 200 times higher thus resulting in a value of $B(\nu )$ 
higher than estimated above.   For this region the calculation done in section 3.2 is indicative 
only and it can be better focussed after we have a better understanding of the sources of 
gravitational waves. 

We now consider how our results compare with the standard model.  

\section{A COMPARISON WITH STANDARD COSMOLOGY}

In the standard model inflation generates both (scalar) density perturbations and (tensor) gravity wave perturbations that are predicted to evolve independently, with uncorrelated power spectra.  The amplitudes of  tensor modes fall off rapidly on sub-Hubble radius scales. The tensor modes on the scales of Hubble-radius along the line of sight to the last scattering distort the photon propagation and generate an additional anisotropy pattern predominantly on the largest angular
scales.  

On large angular scales, the curl component (B-mode)  of CMB polarisation is a unique signature of inflationary gravitational waves.  The amplitude of B-mode CMB polarisation is a direct probe of the energy scale of early universe physics that generates the primordial metric perturbations.  The relative amplitude of tensor to scalar perturbations, $r$, sets the energy scale for inflation $E =
3.4\times 10^{16}$~GeV~$r^{1/4}$.  A measurement of $B$--mode polarisation on large scales would give us this amplitude, and hence {\it a direct determination of the energy scale of inflation.}  The spectrum of stochastic gravitational wave energy density spans a vast range from the cosmological Hubble scales down to scale of centimetres (dictated by the energy scale of reheating). 

In contrast to the above expectation, which depends of course on the type of inflationary past, in the QSSC discussed in this paper we do not expect a strong signal for polarisation in GWB to survive since the different MCEs are randomly oriented. So far as the intensity of GWB is concerned, an order of magnitude estimate given in $\S$3.2 above may be compared with standard cosmology 
where inflation leads to a flat spectrum of background at $\Omega_{\rm GWB} \sim 10^{-14}$.   The spectral form $\alpha \nu^{-1}$ may be another distinguishing feature of the QSSC.  

\section{CONCLUSION}

Although still well below detectable limits, the gravitational wave background in the QSSC presents a coherent answer that can be eventually tested. Since the MCEs are expected to be randomly oriented, we do not expect a strong polarisation signal to emerge. The spectral signal over limited frequencies will be like ${\nu}^{-n}$ with $n$ between 1 and 2.  This signal and the lack of polarisation may be looked for as indicators of QSSC, whereas a clear signal highlighting polarisation and spectral features characteristic of the standard model will go in its support. For the time being, however, we have to be patient and look for improvements in the detection techniques.  The same applies to the practical use of formula (34) for individual sources.  

\section{Acknowledgements}

We thank Tarun Souradeep for illuminating discussions on GWB in the standard cosmology. We further thank Sanjit Mitra for useful discussions on the manuscript and Drs Ken Roberts and John Drozd both from UWO for useful suggestions on the  analytical calculations.  This paper was completed when one of us (JVN) was visiting the Perimeter Institute (PI) in Waterloo, Canada. He thanks the PI for warm hospitality. SRV thanks IUCAA for gracious hospitality and Kings University College (UWO) for its steady support in his research endeavours. 

This research was supported in part by Perimeter Institute for Theoretical Physics. Research
at Perimeter Institute is supported by the Government of Canada through Industry Canada and
by the Province of Ontario through the Ministry of Research and Innovation.

\section*{References}

\noindent Bondi H., Gold T., 1948, MNRAS, 108, 252 \\

\noindent Hild et al, 2011, Class. Quant. Grav., 28, 094013 \\

\noindent Hoyle F., 1948, MNRAS, 108, 372 \\

\noindent Hoyle F., Burbidge G., Narlikar J.V., 1993, ApJ, 410, 437 \\

\noindent Hoyle F., Burbidge G., Narlikar J.V., 1994a, MNRAS, 267, 1007 \\

\noindent Hoyle F., Burbidge G., Narlikar J.V., 1994b, A\&A, 289, 729 \\

\noindent Hoyle F., Burbidge G., Narlikar J.V., 2000, A Different Approach to Cosmology, 
Cambridge Univ. Press, Cambridge \\ 

\noindent Hoyle F., Narlikar J.V., 1962, Proc. R. Soc. A, 270, 334 \\

\noindent Hoyle F., Narlikar J.V., 1963, Proc. R. Soc. A, 273, 1 \\

\noindent Hoyle F., Narlikar J.V., 1964a, Proc. R. Soc. A, 278, 465 \\

\noindent Hoyle F., Narlikar J.V., 1964b, Proc. R. Soc. A, 282, 178 \\

\noindent Hoyle F., Narlikar J.V., 1966, Proc. R. Soc. A, 290, 143 \\

\noindent Lewin L., Polylogarithms and Associated Functions, North Holland, New York 1981.\\

\noindent Lopez-Corredoira, M., 2014, ApJ, 781, 96 \\

\noindent Marassi S., Ciolfi R., Schneider R., Stella L. and Ferrari V., 2011, MNRAS, 411, 2549 \\

\noindent  Molli M., Venkataramaniah K., Valluri S.R., 2011, Can. J.Phys, 89, 1171-1178\\

\noindent Narlikar J.V., 2002, An Introduction to Cosmology, Cambridge Univ. Press, Cambridge \\

\noindent Narlikar J.V., Vishwakarma R.G., Burbidge G., 2002, PASP, 114, 1092 \\

\noindent Narlikar J.V., Vishwakarma R.G., Hajian Amir, Souradeep Tarun, Burbidge G.,  Hoyle F., 2003, ApJ, 585, 1 \\ 

\noindent Prudnikov A.P., Brychkov Yu. A., Marichev O.I., Integrals and Series, Gordon and Breach, New York, Vol. 1, 1986 \\ 

\noindent Sachs R., Narlikar J.V., Hoyle F., 1996, A\&A, 313, 703 \\

\noindent Sami M., Toporensky Alexey, 2004, Mod. Phys. Lett. A, 19, 1509 \\

\noindent Sarmah B.P., Banerjee S.K., Dhurandhar S.V., Narlikar J.V., 2006, MNRAS, 369, 89 (Paper I) \\ 

\noindent Sathyaprakash B.S., Schutz B.F., 2009, Living Rev. in Relativity, 12, 2 \\ 

\noindent Vishwakarma R.G., Narlikar J.V. , 2010, RAA, 10, 12, 1195 \\

\end{document}